\documentclass[12pt]{article}
\usepackage{amsmath}
\usepackage{graphicx}
\usepackage{natbib}
\usepackage{url} % not crucial - just used below for the URL 

\newcommand{\blind}{0}

\usepackage{multirow}
\usepackage{subfigure}
\usepackage{makecell}
\usepackage{booktabs}
\usepackage{array}
\usepackage{url}
\usepackage{algorithm}
\usepackage{algorithmic}
\usepackage{bm}
\usepackage{fullpage}
\usepackage{wrapfig}
\usepackage{lipsum}
\usepackage{mathrsfs}
\usepackage{dsfont}
\usepackage{titling}
\usepackage{mathtools}

\usepackage[colorlinks,
linkcolor=red,
anchorcolor=blue,
citecolor=blue
]{hyperref}

\usepackage{star}

\newcommand \rw{\mathrm{w}}

\def\T{\mathrm{\scriptstyle T}} %%%transpose operator

\newcommand {\V}{\mathrm{V}}

\newcommand{\Rom}[1]{\text{\uppercase\expandafter{\romannumeral #1\relax}}}

\usepackage{mathtools}

\DeclarePairedDelimiter\floor{\lfloor}{\rfloor}

\def\##1\#{\begin{align}#1\end{align}}
\def\$#1\${\begin{align*}#1\end{align*}}

\renewcommand{\baselinestretch}{1.1}

\usepackage{enumitem}
\usepackage[toc,page]{appendix}

\PassOptionsToPackage{hyphens}{url}\usepackage{hyperref}
\hypersetup{
    colorlinks=true,% make the links colored
}

\newcommand{\RN}[1]{%
  \textup{\uppercase\expandafter{\romannumeral#1}}%
}

\usepackage{bbm}
\usepackage{xcolor}
\usepackage{float}

\usepackage[justification=centering]{caption}
\usepackage{setspace}

\newcommand\s{\mathrm{S}}
\newcommand\U{\mathrm{U}}

\newcommand{\IRSS}{I_{\textnormal{RSS}}}
\newcommand\W{\mathrm{W}}

\usepackage{mathtools}

\usepackage{booktabs}
\usepackage[flushleft]{threeparttable}

\usepackage{graphics}

\usepackage{xr}
\makeatletter
\newcommand*{\addFileDependency}[1]{% argument=file name and extension
  \typeout{(#1)}
  \@addtofilelist{#1}
  \IfFileExists{#1}{}{\typeout{No file #1.}}
}
\makeatother

\usepackage[toc,page]{appendix}

\usepackage{siunitx} %

\usepackage{adjustbox}
\usepackage{sectsty}
\subsectionfont{\normalfont\bfseries\itshape}

\usepackage{comment}

\usepackage{authblk}
\usepackage{tabularx}

\newcommand\clearrow{\global\let\rowmac\relax}
\clearrow

\usepackage{color}

\begin{document}

\def\spacingset#1{\renewcommand{\baselinestretch}%
{#1}\small\normalsize} \spacingset{1}

%%%%%%%%%%%%%%%%%%%%%%%%%%%%%%%%%%%%%%%%%%%%%%%%%%%%%%%%%%%%%%%%%%%%%%%%%%%%%%

\if0\blind
{
  \title{\bf Supervised Principal Component Regression for Functional Responses with High Dimensional Predictors}
  \date{}
  \author[1]{Xinyi Zhang}
  \author[2]{Qiang Sun} 
  \author[2]{Dehan Kong}
  \affil[1]{School of Medicine, Johns Hopkins University, Baltimore, Maryland, USA}
  \affil[2]{Department of Statistical Sciences, University of Toronto, Toronto, Ontario, Canada}
  \maketitle
  
} \fi

\if1\blind
{
  \bigskip
  \bigskip
  \bigskip
  \begin{center}
    {\LARGE\bf Title}
\end{center}
  \medskip
} \fi

\bigskip
\begin{abstract}
We propose a supervised principal component regression method for relating functional responses with high dimensional predictors. Unlike the conventional principal component analysis, the proposed method builds on a newly defined expected integrated residual sum of squares, which directly makes use of the association between the functional response and the predictors. Minimizing the integrated residual sum of squares gives the supervised principal components, which is equivalent to solving a sequence of nonconvex generalized Rayleigh quotient optimization problems. We reformulate the nonconvex optimization problems into a simultaneous linear regression with a sparse penalty to deal with high dimensional predictors. Theoretically, we show that the reformulated regression problem can recover the same supervised principal subspace under certain conditions. Statistically, we establish non-asymptotic error bounds for the proposed estimators when the covariate covariance is bandable. We demonstrate the advantages of the proposed method through numerical experiments and an application to the Human Connectome Project fMRI data.
\end{abstract}

\noindent%
\vfill

\newpage
\spacingset{1.75} % DON'T change the spacing!

\section{Introduction}\label{sec:intro}
Functional magnetic resonance imaging (fMRI) data have provided researchers with unprecedented insights into the inner workings of the human brain,  brain connectivity, and predictions about psychological or disease propagation. This paper studies  how the clinical variables are associated with the blood-oxygen-level-dependent (BOLD) signals in a functional region over a period of time, in which BOLD signals can be characterized as a single functional curve and are often modeled as a smooth function. This association can then be used to make predictions and inferences about psychological or disease states. Despite fMRI data being the focus of many recent scientific studies \citep{Preti2016}, the association between the clinical measurements and the BOLD signals has not yet been well understood, possibly due to the high dimensionality of clinical measures collected in a study.

A functional-on-scalar linear model is commonly used to assess the relationship between a functional response and predictors \citep{fara}. To handle high dimensional predictors, recent research has explored penalized estimation methods, such as those discussed in \citet{Chen2016} and \citet{Barber2017}. Typically, these methods first obtain a low dimensional representation of the functional response using techniques like functional principal component analysis \citep{Yao&Muller2005}, and then examine the relationship between the scalar covariates and this low-dimensional representation. The primary objective is to estimate the effect of each scalar covariate on the functional response.

Dimension reduction is particularly important in high dimensional settings as it helps reduce the data complexity.
The Lasso penalized regression is one such method that can force the effect estimation of uninformative covariates to be exactly zero, However, it fails to deal with strong collinearity among predictors. Another common approach is principal component analysis, which aims to extract a lower-dimensional subspace that captures most of the variation in the covariates only. Nonetheless, classical principal component analysis does not utilize the response and thus may miss the subspace which contains the most predictive information.

Within the context of functional-on-scalar linear model, in this paper we propose a supervised principal component regression method that directly incorporates the association information between the response and the predictors. Our development leverages a new notion of the expected integrated residual sum of squares to account for the functional nature of the response. Computationally, the proposed method can be formulated as a sequence of sparse generalized Rayleigh quotient problems with orthogonality constraints. Due to the nonconvexity, solving such problem is often computationally intractable.  \citet{Witten2011} proposed a majorize-minimization algorithm for a related problem, which is the linear discriminant analysis, in a sequential manner. Their algorithm generally has slow computational convergence unless the within-class covariance is assumed to be isotropic.  Moreover, the global convergence to the optimal solution cannot be guaranteed generally. Recent advances have also been made in solving nonconvex sparse generalized Rayleigh quotient problems \citep[e.g.][]{ tan2018sparse, Guan2022}. However, these methods either only compute the leading projection direction, or still proceeds in a sequential fashion, which may entail considerable computational costs, especially for large dimensions. Additionally, a sequential procedure possibly lead to error propagation and estimation results may be numerically unstable. 
To alleviate such computational burdens, we reformulate the proposed method into a convex penalized linear regression problem, which can solve multiple supervised principal components simultaneously and further enables much faster computation.

There are also some related work on supervised principal component analysis but within different contexts. \citet{Bair2006} considered a scalar response and proposed a two-step procedure that first selects a subset of the predictors based on their associations with the outcome and then applies the classical principal component analysis to those selected variables. Their approach does not fully utilize the associations between the response and the predictors because the association information is only leveraged to select a subgroup of features but not to estimate principal components. More recently, \citet{Li2016} studied a similar problem and formulated their approach as a special case of latent variable models, which requires a sequential and iterative estimation procedure. As previously discussed, sequential algorithms are computationally expensive and estimation errors are likely to be propagated for these procedures.

Our contributions are three-fold. First, we propose a  supervised principal component analysis with high dimensional predictors and functional response, which has not yet been studied in the literature to the best of our knowledge. Also, all existing methods on supervised principal component analysis only investigated the setting where the dimension of predictors is smaller than the sample size, while we allow the dimensionality to grow at an exponential order of the sample size. Second, we reformulate our method into a  simultaneous regression problem by exploiting the new concept of integrated residual sum of squares for functional data, which improves computational efficiency. Third, we show that the reformulated problem can recover the same subspace. Motivated by our fMRI data application, specific to the bandable covariance structure, we further establish non-asymptotic error bounds for the proposed estimators.

The rest of this article is organized as follows. In Section \ref{s2}, we introduce our method with motivation and reformulate it into a penalized multivariate regression problem, followed by the detailed estimation procedure.  
Section \ref{s3} is devoted to theoretical results. Numerical experiments are carried out in Section \ref{s4}. In Section \ref{s5}, we apply our method to the cortical surface emotion task-related fMRI data from Human Connectome Project dataset. We close with a discussion on possible future works in Section \ref{s6}.

Throughout, for a matrix $A\in \mathbb{R}^{n\times m}$, $\tr(A)$ represents the trace. For $q\in (0,\infty]$, $\|A \|_q=\sup_{x \in \mathbb{R}^m\neq 0 } \|Ax \|_q/\| x\|_q$ denotes the matrix $ q $ norm. 
Specifically, $\|A\|_{1}=\mathrm{max}_{j}\sum_{i=1}^{n} |A_{ij}|$ and $\|A\|_{\infty}=\mathrm{max}_{i}\sum_{j=1}^{m} |A_{ij}|$. For $q_{1},q_{2}\in [0,\infty]$, 
$\|{A} \|_{q_{1},q_{2}}=\|(\|{A}_{\cdot 1}\|_{q_{2}}, \|{A}_{\cdot 2}\|_{q_{2}}, \cdots, \|{A}_{\cdot m}\|_{q_{2}})\|_{q_{1}}$ is the pseudonorm of $A$. Specifically, $\| A\|_{\infty, \infty}$ refers to the maximal element of $A$ in absolute value. $\|A\|_{\mathrm{F}}$ represents the Frobenius norm.
We write $f(t) \gtrsim g(t)$, if there exists a constant $ C>0$ such that $f(t)\geq Cg(t)$, and  $f(t) \lesssim g(t)$ if $f(t)\leq Cg(t)$. For  probability measure $Q$ on a measurable space $(D, \mathcal{D})$, define $Qf:=\int f dQ$, and $\mathcal{L}^p(Q)$ denotes the space of all measurable functions $f: D\rightarrow \mathbb{R}$ such that $\| f\|_{Q,p}:=(Q|f|^p)^{1/p}<\infty$ for $p\in [1,\infty]$.

\section{Motivation and Methodology}\label{s2}

\subsection{Supervised principal component regression with functional responses}\label{motivation}
Let $\{Y(t):t\in T\}$ be a centered functional response with $ T \subset \mathbb{R}$ a compact support, and $X=(X_1, \ldots, X_p)^\T \in \mathbb{R}^{p}$ represents centered covariates. We consider the following functional-on-scalar linear model
\begin{equation}\label{e1}
Y(t) = X^\T\beta(t)+\epsilon(t),
\end{equation}
where $\beta(t)=(\beta_1(t), \ldots, \beta_p(t))^{\T} $ denotes the functional coefficients and $ \epsilon(t) $ is random error with zero mean and finite second moment. Further assume that $\epsilon(t)$ is independent of $X$.

To deal with high dimensional predictors, we first  obtain a low-dimensional representation of $X \in \mathbb{R}^p$. Then we characterize the relationship between this low-dimensional vector of new predictors and the response. One conventional way is the principal component regression  \citep{Hotelling1957}, which searches for linear combinations of the original covariates that preserve as much variability as possible, 
and then regresses the response on the new covariates.
However, the method ignores the response variable and thus can only discover a sequence of directions that explain the maximum variation of the covariates. Instead, we develop a supervised principal component regression method, which finds the low dimensional representation of the covariates by embracing the response information.

To fix the idea, we start with the  warm-up case where $p < n$. The projection of $X$ onto direction $\rw_1 \in \mathbb{R}^p$ is $ X^\T \rw_1 $. 
Suppose the functional response $ Y(t) $ and the projection $ X^\T \rw_1 $ are linearly associated. The optimal regression function $ \gamma_{\rw_1}^*(t) $ regarding $ X^\T \rw_1 $ minimizes the expected integrated residual sum of squares defined as
$
\IRSS\{\gamma(t)\}=\int_T E[\{Y(t)-X^\T \rw_1\gamma(t)\}^\T \{Y(t)-X^\T \rw_1\gamma(t)\}]{{\rm d}t}.
$
It is easy to show that $\gamma_{\rw_1}^*(t) $ has a closed form $\gamma_{\rw_1}^*(t)=\{\rw_1^{\T} E(X X^\T ) \rw_1\}^{-1}\rw_1^{\T} E\{X Y(t)\}$. Plugging $\gamma_{\rw_1}^*(t) $ into the ${\IRSS}$ formula yields 
\begin{equation*}
{\IRSS}(\gamma_{\rw_1}^*)=\int_T (E\{Y^2(t)\}-[E\{X Y(t)\}]^\T \rw_1\{\rw_1^{\T} E(X  X^\T ) \rw_1\}^{-1}\rw_1^{\T}[E\{X Y(t)\}]) {{\rm d}t}, 
\end{equation*}
which is a function of $ \rw_1 $. The optimal direction 
among all possible $ \rw_1 $s is given in Proposition \ref{pp1} below.

\begin{proposition}\label{pp1}
If $ \rw_{1,0} $ is a minimizer of $ {\IRSS}(\gamma_{\rw_1}^*) $, then $ \rw_{1,0}$ satisfies 
\begin{equation}\label{e3}
\rw_{1,0}=\arg \max_{\rw_1} \frac{\rw_1^{\T} \Sigma_{xy} \rw_1}{\rw_1^{\T} \Sigma_x \rw_1}, 
\end{equation}
where $ \Sigma_x=E(XX^\T) \in \mathbb{R}^{p\times p}$ and $ \Sigma_{xy}=\int_T E\{X Y(t)\}[E\{X Y(t)\}]^\T {{\rm d}t} \in \mathbb{R}^{p \times p}$. 
\end{proposition}

One can see that  $c\rw_{1,0}$ for any constant $c>0$ is also a solution to the optimization problem \eqref{e3}. 
We then restrict $\rw_1^{\T}\Sigma_x{\rw_1}=1$ and consider the problem below
\begin{equation}\label{e4}
\rw_{1}^*=\arg \max_{{\rw}_1} {\rw_1}^\T\Sigma_{xy} {\rw_1}~~\textnormal {subject~to~} \rw_1^{\T}\Sigma_x{\rw_1}=1. 
\end{equation}
The vector $\rw_{1}^* \in \mathbb{R}^p$ denotes the first supervised principal component direction, which is unique up to sign flips. Analogously, we can obtain the other $K-1$ supervised principal component directions for $K < p$
by sequentially solving the problems below 
\begin{eqnarray}\label{e5}
&\rw_{k}^*=\arg\max_{{\rw_k}}  \rw_k^{\T}\Sigma_{xy} {\rw_k}\textnormal {~subject~to~}  \rw_k^{\T}\Sigma_x{\rw_k}=1, \rw_k^{\T}\Sigma_x{\rw_{j}^*}=0, 1\leq j<k,
\end{eqnarray}
for $1\leq k \leq K$. Denote $ \mathrm{W}^*=(\rw_{1}^*, \ldots, \rw_{K}^*) \in \mathbb{R}^{p\times K}$ the top $ K $ supervised principal component directions.

In the high dimensional setting where $p>n$, it is natural to introduce sparsity constraint to the supervised principal component directions. To this end, we estimate a sparse $\W^*$ via 
\begin{align}\label{optW_sparse}
        &\rw_{k}^*=\arg\max_{{\rw_k}}  \rw_k^{\T}\Sigma_{xy} {\rw_k} \nonumber\\ 
        & \textnormal {~subject~to~}  \rw_k^{\T}\Sigma_x{\rw_k}=1, \rw_k^{\T}\Sigma_x{\rw_{j}^*}=0, \|\rw_{k}\|_q \leq t
\end{align}
for $1\leq j < k$ and $1 \leq k \leq K$. 
\begin{remark}
The canonical correlation analysis (CCA) might be another option to explore the relationship between two data sets. However, CCA and our approach have fundamental differences. CCA focuses on finding linear combinations of both covariates $ X$ and the response $Y(t)$ that are highly correlated. Instead, our approach aims to minimize the integrated residual sum of squares by keeping the original response and seeking a new predictor that is a projection of $X$ onto a low-dimensional subspace. 
\end{remark}

\subsection{Convex reformulation}\label{convex_reform}

There are some existing works \citep[e.g.][]{tan2018sparse, Guan2022} along the line of solving the non-convex problem \eqref{optW_sparse} for $q=0$ or $q=1$, either with a focus on the leading direction only or estimate multiple directions in a sequential manner. 
Additionally, statistical properties for these methods have only been derived for the leading projection direction. 
Here we instead propose to estimate multiple supervised principal component directions simultaneously through a convex relaxation. Our method is computationally faster, especially when $p$ is large. 
We later establish theoretical guarantees for estimating multiple directions, and empirically demonstrate the computational gain using simulations.

To motivate, we first focus on the problem \eqref{e5} without the sparsity constraint.  
Let $\{(\lambda_{j},\eta_{j})\}_{j=1}^p$ be eigenpairs of $\Sigma_{xy}$, where 
$\lambda_1\geq \lambda_2 \geq \ldots \geq \lambda_p\geq 0$.
Decompose the matrix ${\Sigma}_{xy}$ into 
${\Sigma}_{xy}=\U D \U^{\T}+R$, where ${\U}\in \mathbb{R}^{p\times K}$ is formed by the  
first $K$ eigenvectors of $ \Sigma_{xy}$ and $D$ is a diagonal matrix with diagonal elements equal to $ \lambda_1, \ldots, \lambda_K$. It can be shown that the following convex optimization problem 
\begin{equation}\label{e7}
\V^{*} = \argmin_{\V\in \mathbb{R}^{p\times K}} \frac{1}{2}\|\Sigma_{x}^{-1/2} \U-\Sigma_{x}^{1/2} \V \|_{\mathrm{F}}^{2}. 
\end{equation}
can recover the same principal subspace obtained from the original nonconvex Rayleigh quotients problem \eqref{e5} under the low-rank assumption imposed on $\Sigma_{xy}$. We formalize this result in Theorem \ref{thm1}. 

\begin{theorem}[Equivalence]\label{thm1} 
Assume $\Sigma_{xy}$ is of low-rank. We have $\text{span}\{\V^*\} = \text{span}\{\mathrm{W}^{*}\}$, which are the linear subspaces generated by the columns of $\V^*$ and $\mathrm{W}^{*} $ respectively.
\end{theorem}

\begin{remark}
In practice, $\Sigma_{xy}$ may not be of exact low rank, however, it is often nearly low-rank \citep{negahban2011estimation}. Accordingly, solution to the  convex relaxation \eqref{e7} can approximate solution to the original nonconvex optimization problem \eqref{e5}.
\end{remark}

Motivated by the equivalence between  $\W^*$ and $\V^*$ as shown in Theorem \ref{thm1}, with high dimensional predictors, we estimate projection directions $ \W^*$ via
\begin{equation}\label{optV_penalize}
        \V^* = \argmin_{\V \in \mathbb{R}^{p \times K}} \frac{1}{2}\|\Sigma_{x}^{-1/2} \U-\Sigma_{x}^{1/2} \V \|_{\mathrm{F}}^{2} + \lambda \| \V\|_{1,1}.
\end{equation}
We believe that the span of $ \W^*$ would be close to the span of $\V^*$. 
See \citet{mai2012direct} for a similar methodology motivation in high dimensional linear discriminant analysis.

\begin{remark}\label{remark_altern}
Alternatively, one can estimate the subspace $\V$  by first identifying the eigenvectors of $ \Sigma_x^{-1/2} \Sigma_{xy} \Sigma_x^{-1/2} $, say $\s$, then solving a $\ell_1$ regularized regression problem with response $\s$ and predictor $\Sigma_x^{1/2}$. This approach is referred to as ``SPCR-A". Establishing theoretical guarantees for ``SPCR-A" can be challenging, as we need to control estimated eigenvectors of 
$ \Sigma_x^{-1/2} \Sigma_{xy} \Sigma_x^{-1/2} $. We thus focus on proposal \eqref{optV_penalize} in this article. 
Discussion of ``SPCR-A" and empirical comparison are included in Section S6 of the supplement. 
\end{remark}
In practice, the optimization problem \eqref{optV_penalize} involves unknown $ \Sigma_x$ and $\U$. 
Their empirical estimations are described in Section \ref{estimation} below.

\subsection{Estimation of supervised principal components}\label{estimation}
Suppose that observation $ \{(X_i, Y_i(t)) \}_{i=1}^{n}$ are independently and identically generated from $ (X, Y(t)) $.
Let $ \mathbb{X}= (X_1, \ldots, X_n)^\T \in \mathbb{R}^{n\times p}$ be the design matrix and $ \mathbb{Y}(t)=(Y_1(t), \ldots, Y_n(t))^\T $ denotes an $ n $-dimensional functional vector. 

\textit{Estimation of $\Sigma_x$.}
In high dimensions, the sample covariance matrix $ \widehat\Sigma_x^0= n^{-1}\mathbb{X}^\T\mathbb{X}$  is a poor estimator of $\Sigma_x$ as it is rank-deficient and the
the associated eigenvalues and eigenvectors can be far away from those of $\Sigma_x$. Motivated by various scientific applications, regularization-based methods have been proposed to estimate $\Sigma_x$, such as banding, tapering, and thresholding. See \cite{Cai2016} for a complete review. 
Driven by our data application where predictors have been grouped into different categories, we restrict our analysis to the bandable covariance class, in which correlation decays between covariates far apart in the ordering. Moreover, we adopt the banding estimator proposed in \citet{BICKEL2008}.
Explicitly, the banded covariance estimator takes the form of
$
\widehat{\Sigma}_x = B_b(\widehat\Sigma_x^0)=[(\widehat\Sigma_x^0)_{ij}\mathbbm{1}(|i-j|\leq b)].
$
The bandwidth $b$ is selected by the random sampling approach introduced in \cite{BICKEL2008}. 
Specifically, we randomly split the sample into two halves $H_1$ and $H_2$ of sizes $n_1$ and $n_2$ respectively. We obtain the banded estimate from $H_1$ and use the sample covariance matrix of $H_2$ as the ``true" covariance matrix to choose the best $b$. 
In numerical studies, we take 
$n_1=\floor{n/3}$ as recommended in \cite{BICKEL2008}. Let $ \widehat{\Sigma}_1^{(s)}$ and $ \widehat{\Sigma}_2^{(s)}$ denote the sample covariance matrices for $H_1$ and $H_2$ from the $s$th split, for $s= 1, \ldots, S$. We set $S=20$. Then we calculate the estimation error over $S$ splits by 
$
\widehat{R}(b) = \frac{1}{S}  \sum_{s=1}^S \| B_b(\widehat{\Sigma}_1^{(s)}) - \widehat{\Sigma}_2^{(s)}\|_{1,1}, 
$
and the optimal $b$ is selected as
$
\widehat{b} = \argmin_b{\widehat{R}(b)}
$ based on grid search.

\textit{Estimation of $U$.}  
We first consider the case where the functional response trajectory can be fully observed.
The cross-covariance $ \Sigma_{xy} $ is estimated by  $ \widehat\Sigma_{xy}=n^{-2}\int_T \mathbb{X}^\T\mathbb{Y}(t)\mathbb{Y}(t)^\T \mathbb{X} {{\rm d}t} $. 
The eigenpairs of $\widehat\Sigma_{xy}$ are denoted by $\{(\widehat\lambda_{j},\widehat{\eta}_{j})\}_{j=1}^p$, where $\widehat\lambda_{j}$'s are in the decreasing order. Similar to the decomposition of $\Sigma_{xy}$, we can write 
$ \widehat{\Sigma}_{xy}=\widehat \U \widehat{D} {\widehat \U}^\T+\widehat{R}$, where ${\widehat \U}\in \mathbb{R}^{p\times K}$ consists of the first $K$ eigenvectors of $\widehat{\Sigma}_{xy}$ and $\widehat{D}$ is a diagonal matrix with diagonal elements 
$\widehat{\lambda}_1,\ldots, \widehat{\lambda}_K$.

In real applications, the functional responses are often measured at a dense grid of regularly spaced time points with measurement errors. 
Specifically, rather than observing the entire response trajectory $ Y_i(t)$, we only have access to intermittent noisy measurements $ W_{il}= Y_i(t_{il})+ \epsilon_{il}$, where $\{t_{il}, i=1, \ldots, n; l=1, \ldots, L\}$ are the time points and $\{\epsilon_{il}, i=1, \ldots, n, l=1, \ldots, L\}$ are independently and identically distributed measurement errors and moreover are independent of $Y_i(t)$. In this case, we apply the curve-by-curve smoothing \citep{li2010generalized, Kong2016} as a pre-processing step and then estimate $\Sigma_{xy}$ using the smoothed curve. In particular, for each individual curve, we use smoothing spline based on noisy observations $\{W_{il}\}_{l=1}^{L}$, to obtain the smoothed curve $\tilde{Y}_i(t)$ and their realization $\tilde{Y}_i(t_{il})$ at the grid points. 
The estimation procedure for fully observed functional responses can be readily adapted for this scenario: Estimate $\Sigma_{xy}$ by $\widehat\Sigma_{xy}=\int_T \mathbb{X}^\T\tilde{\mathbb{Y}}(t)\tilde{\mathbb{Y}}(t)^\T \mathbb{X} {{\rm d}t}/n^2$ with $\tilde{\mathbb{Y}}(t) = \big(\tilde{Y}_1(t), \ldots, \tilde{Y}_n(t) \big)^{\T} $, and all other steps remain unchanged. If the functional responses are irregularly or sparsely sampled, one may apply the smoothing techniques proposed in \citet{Yao&Muller2005} to obtain the smoothed curve $\tilde{Y}_i(t)$ and estimate $ \Sigma_{xy} , U$ accordingly.

Given empirical estimates of $\Sigma_x, U$, the supervised principal component directions can be estimated simultaneously via the following convex Lasso problem
\begin{align}\label{e9}
\widehat \V = \underset{\V}{\argmin} \frac{1}{2}\|\widehat{\Sigma}_{x}^{-1/2} \widehat{\U}-\widehat{\Sigma}_{x}^{1/2} \V \|_{\mathrm{F}}^{2}+\lambda\| \V\|_{1,1},
\end{align}
from which the supervised principal components are constructed. 
In this work, we use the \texttt{R} package \texttt{glmnet}  for implementation. To select optimal tuning parameters $ (K^*, \lambda^*)$, we use a two-dimensional grid search based on 5-fold cross-validation.

\subsection{Estimation of functional coefficients}
Before proceeding to the supervised principal component regression estimate of the functional coefficients $\beta(t)$ in the underlying model \eqref{e1}, we elucidate the connection of the supervised principal components and the model \eqref{e1}. Let $\V^{\T}  X \in \mathbb{R}^{K}$ denote the low-dimensional representation of the high-dimensional predictor $ X\in \mathbb{R}^p$, where $ \V \in \mathbb{R}^{p \times K}$ for $K \ll p$ represents the supervised principal component directions. The underlying model \eqref{e1} can be rewritten in terms of $\V^{\T}  X$ as follows
\begin{equation}\label{mod_reformV}
Y(t) = (X^{\T} \V) \gamma(t) + \tilde{\epsilon}(t)+\epsilon(t), 
\end{equation}
where $\gamma(t)\in \mathbb{R}^K $ is the coefficient corresponding to the low-dimensional projection $ \V^{\T}  X $, $ \epsilon(t)$ is the random error in the true model \eqref{e1}, and $ \tilde{\epsilon}(t)$ characterizes the approximation error $X^{\T}\beta(t) - (X^{\T} \V) \gamma(t)$. The model \eqref{mod_reformV} can be regarded as a functional-on-scalar linear model with covariates $\V^{\T}  X$, the coefficient $ \gamma(t) $, and the random error $\tilde{\epsilon}(t)+\epsilon(t)$.

Given the estimated supervised principal component directions $ \widehat{\V} \in \mathbb{R}^{p \times K^*}$ obtained from the proposed procedure, the regression coefficient estimate $ \widehat{\gamma}(t) $, a $K^*$ dimensional vector, can be calculated by regressing $Y(t)$ on $ \widehat{\V}^{\T} X$. 
If one is interested in estimating the coefficient function $ \beta(t) $ in model \eqref{e1}, 
then based on the connection between models \eqref{e1} and \eqref{mod_reformV}, the supervised principal component regression estimate of $ \beta(t)$ is given by $ \widehat{\beta}^{(K^*)}(t)= \widehat{\V}\widehat{\gamma}(t) $. Our proposed estimation procedure shares the same spirit as the conventional principal component regression \citep{Hotelling1957}.

\begin{remark}\label{multiY}
The proposed approach can also be applied to the multivariate response linear regression model, where the response is a $T$ dimensional vector $ {\bf Y}=(Y_1, \ldots, Y_T)^{\T}$. We can obtain at most $ T $ supervised principal component directions in this case. 
\end{remark}

\section{Theoretical Properties}\label{s3}
In this section, we provide theoretical understanding of the proposed method. Assuming that the functional response trajectory is fully observed, we establish error bound for the estimated subspace. Let $\mathcal{S}$ be the support set of $\V^{*}$ of size $s$, i.e. $\mathcal{S}=\{(i,j): \V^{*}_{ij}\neq 0 \}$. 
We restrict our analysis to 
the class of bandable covariance matrices \citep{BICKEL2008} defined as
\begin{align*}
\mathcal{U}(\epsilon_0, \alpha, C)=\bigg\{\Sigma: \max_j \sum_{i}&\{  |\sigma_{ij}|:|i-j|>b  \}\leq Cb^{-\alpha} \text{ for all } b>0, \nonumber \\
& \text{and } 0<\epsilon_0 \leq \lambda_{\min}(\Sigma)\leq \lambda_{\max}(\Sigma)\leq 1/\epsilon_0 \bigg\}
\end{align*}
for some  positive constants $ C $ and $\alpha $.
The assumptions imposed to establish the non-asymptotic error bound are presented below
\begin{itemize} 
\item[(A1)] (Distribution) Assume that $X_{j}$ has mean 0 and variance $ \sigma_{jj}$ and $X_{j}/\sqrt{\sigma_{jj}}$ is Sub-Gaussian with variance proxy $\sigma^2$ for $ j=1, \ldots, p$.  
The covariance matrix $\Sigma_x \in \mathcal{U}(\epsilon_0, \alpha, C)$, with banding parameter $b$ satisfying $b \asymp (\log p/n)^{-1/(2(\alpha+1))}$. Further assume random error $\epsilon(t)$ is a Gaussian process, where $t \in T$ and $T$ is a compact set. 
\item[(A2)] (Conditions on random processes) 
Define $f_{t,j}(X_i)\coloneqq Y_i(t)X_{ij}-E\{Y_i(t)X_{ij}\}$ and $\mathcal{F}_j = \{f_{t,j}: t\in T\}$. Suppose there exists a measurable function $\tau_j \in \mathcal{L}^2(Q)$ 
such that $\|f_{t_1,j}(X_i) - f_{t_2,j}(X_i) \|_{Q,2} \leq  \|\tau_j(X_i)\|_{Q,2} |t_1 - t_2|$ for any $t_{1},t_{2}\in T$ and probability measure $Q$. Further assume there exists a nonnegative function $F\in \mathcal{L}^2(P)$ with $P$ denoting the distribution of $X_i$
such that $\max_{1\leq j \leq p}\sup_{f_{t,j}\in \mathcal{F}_j} |f_{t, j}(x)|\leq F(x)$.
\item[(A3)] (Bounded support) Assume $\max_{1\leq j \leq p} \sup_{t\in T} |E\{Y(t)X_{j}\}|\leq M$ for some  constant $M>0$ independent of  $ j , t$.
Also, $\max_{1\leq j , l \leq p}|\int_{T} \beta_j(t) \beta_l(t) {{\rm d}t}| < \infty$. 
\item[(A4)] (Coherence and eigengap) Suppose $\mu=(p/K) \max_{1\leq l\leq p}\sum_{j=1}^{K} \eta_{jl}^2$ is bounded by a constant, where
$\eta_{jl}$ denotes the $l$th element of the eigenvector $\eta_j$ of $\Sigma_{xy}$, and $K$ is fixed.  The leading eigenvalue $\lambda_1 \asymp K^3 p\sqrt{(\log p)/n}$. Assume 
$\lambda_{j} - \lambda_{j+1}>\delta$ for each $j = 1, \ldots, K$ and $\lambda_{K+1} = 0$,  and
there exists a constant $\tilde{C}$ such that  $\min\{\lambda_K, \delta \} \geq \tilde{C} K^3 p\sqrt{(\log p)/n}$.
\item[(A5)] (Sparsity) Elements of $\V^*$ defined in \eqref{e7} have bounded support. The level of sparsity for $\V^*$ satisfies $s=o\big( (n / \log p)^{1/3}\big)$. 
\end{itemize}
Next, we define the restricted eigenvalue condition for matrices, which is a direct extension of the restricted eigenvalue condition for vectors \citep{Bickel2009, fan2018lamm}. 
\begin{definition}[Restricted Eigenvalue Condition]\label{def1}
The restricted eigenvalue of the covariance matrix estimate $\widehat{\Sigma}_x$ is defined as 
\begin{align*}
%\kappa_{+}(k',m,\gamma) &=  \sup_{\V} \{\tr(\V^{\T}\widehat{\Sigma}_x \V)/\|\V \|_{1,2}^{2}: \V\in \mathcal{C}(k',m,\gamma)   \}\\
\kappa_{-}(k',m,\gamma) = \inf_{\V} \{\tr(\V^{\T}\widehat{\Sigma}_x \V)/\|\V \|_{1,2}^{2}: \V\in \mathcal{C}(k',m,\gamma)   \},
\end{align*}
where $\mathcal{C}(k',m,\gamma)\equiv \{\V\in \mathbb{R}^{p\times k'}: \mathcal{S}\subseteq \mathcal{E}, |\mathcal{E}|\leq m, \|\V_{\mathcal{E}^c} \|_{1,1}\leq \gamma \|\V_{\mathcal{E}} \|_{1,1}   \}$; $\mathcal{E}=\{(i,j): i \in \{1, \ldots, p\}, j \in \{1, \ldots, k'\}\}$ is a set of two dimensional indices, and $\V_\mathcal{E} \in \mathbb{R}^{p \times k'} $ with $[\V_\mathcal{E}]_{ij}\neq 0$ for $(i,j)\in \mathcal{E}$ and 0 otherwise, and $\V_{\mathcal{E}^c}$ is defined similarly. Then the restricted eigenvalue condition holds if there exist $k',m,\gamma$ such that $\kappa_{-}(k',m,\gamma)\geq \kappa_{*} $, where $ \kappa_{*}$ is a positive constant. 
\end{definition}
We formulate in Theorem \ref{thm2} the non-asymptotic error bound of the estimated subspace. 
The proof is deferred to Section S3 of the supplementary material.

\begin{theorem}\label{thm2}
Suppose (A1)--(A4) and the restricted eigenvalue condition holds with $k'=K, m=2s, \gamma=3$. Let $\lambda \geq 4\max \{ C_{0}(\log p /n )^{1/2} \|\V^{*}\|_{1}, C_0^{\prime} K /\sqrt{p} \}$ for some constants $C_{0}, C_0^{\prime} > 0 $ and further assume $\log p = o(n)$.   
Then with probability going to $1$, we have
\begin{equation}\label{errorBnd_Vhat}
\|\widehat \V-\V^* \|_{1,2} \lesssim \lambda s^{1/2},
\end{equation}
which converges to zero as $n,p \to \infty$ under (A5). 
\end{theorem}
To establish Theorem \ref{thm2}, the main challenges include
showing the convergence rate of $\widehat{\Sigma}_x$ under the max norm and controlling the difference between the estimated eigenvectors of $\widehat\Sigma_{xy}$ and their population counterparts associated with $\Sigma_{xy}$. 
Although we assume $ \Sigma_{xy}$ to be low-rank, this assumption can be relaxed to 
approximate low-rankness, and then similar error bound can be derived with an additional constraint on the low-rank approximation in terms of $\|\Sigma_{xy} - \sum_{j=1}^K \lambda_j \eta_{j}\eta_j^{\T} \|_{\infty}$.

\section{Simulation}\label{s4}

\subsection{Fully observed functional responses}\label{fully_obs}
We first consider functional response whose trajectory is fully observed. Our theories rely on two key assumptions: (1) $\Sigma_{xy}$ is of low-rank and (2) $ \V^* $ is sparse. We compare the performance of our proposal with other methods in two different simulation settings, one with both assumptions satisfied and the other with both of them violated.

\noindent
\textbf{Setting 1}. For $i=1, \ldots, n$, we simulate $ {X}_i $'s independently and identically from multivariate normal distribution with mean $ 0$ and covariance matrix $ \Sigma $, where $ \Sigma_{jj'}=0.25^{|j-j'|} $ for $ 1\leq j,j'\leq p $. Set the support $ T=[0,1] $. Random errors $ \epsilon_i(t) $'s for $i=1, \ldots, n$ are generated independently and identically from a Gaussian process with mean $ 0 $ and covariance function $K(t_1,t_2)=\exp\{-5(t_1-t_2)^2\}$ for $ 0\leq t_1,t_2\leq 1 $. In this simulation setting, we set the true dimension $K_0=3$ by constructing the  functional coefficient $ \beta(t)=\V^*\gamma(t) $, where $\V^*=(\V_{1}^*,\V_{2}^*,\V_{3}^*)^{\T}\in \mathbb{R}^{p\times 3}$ with $\V_{1}^* = (1,1,0,0,\ldots,0,0,0,0)^{\T}, \V_2^* = (0,0,1,1,0,0,\ldots,0,0)^{\T}, \V_3^* = (0,0,0,0,\ldots,0,0,1,1)^{\T}$, and $\gamma(t)=(\gamma_{1}(t),\gamma_{2}(t),\gamma_{3}(t))^{\T}$ with $\gamma_{1}(t) =2\cos(\pi t),\gamma_{2}(t) =3\cos(2\pi t)$,  $\gamma_{3}(t) =5\cos(3\pi t) + 3\sin^{2}(3\pi t)$. Functional responses $ \{Y_i(t)\}_{i=1}^n $ are then generated from model \eqref{e1}. 

\noindent
\textbf{Setting 2}. We set the functional coefficient $ \beta_{j}(t)=[\cos\{\pi t(j+20)/10 \}](15/j^2) $ for $ 1\leq j\leq p $ and all the other settings remain the same as setting 1. 

\noindent
We include the following methods for comparison: 
(i) ``SPCR" represents the proposed method \eqref{e9}. Default candidates for getting optimal
tuning grid parameters $(K,\lambda)$ via 5-fold cross-validation are: $K_{\max}$ is set to be $\min(\text{rank}(\widehat{\Sigma}_{xy}), 30)$; $\lambda$ takes value from an equally spaced sequence of length 50 between 0.005 and 0.2.
(ii)``UPCR" is the conventional principal component regression. The number of principal components is selected by 5-fold cross-validation. 
(iii)``Superpc" refers to the supervised principal component analysis in \cite{Bair2006}. As ``Superpc" only allows for univariate response, we use $
    m_j^2=\sum_{l=1}^L \big\{\sum_{i=1}^n Y_i(t_{l}) X_{ij} \big\}^2
    $
    in their screening step to screen out uninformative predictors for $j =1, \ldots, p$.     
(iv)``CCA" represents the canonical correlation analysis. We adopt the regularization approach in \cite{Witten2009} to obtain sparse canonical vectors. The algorithm is implemented using R package ``PMA". The number of canonical directions is chosen by 5-fold cross-validation. Comparisons with other sparse CCA \citep{Tenenhaus2014, tan2018sparse, Guan2022} can be found in Section S8 of the supplementary material. 
(v)``SPCR-no penalty" is the optimization problem \eqref{e9} without $\ell_1$ penalty. 
(vi)``riﬂe-GEV" represents the truncated Rayleigh flow approach \citep{tan2018sparse} which solves sparse generalized eigenvalue problem with $\ell_0$ penalty. As the algorithm can only solve the leading direction, we set $ K=1$ for this method.
(vii)``PGD-GEV" is the proximal gradient descent algorithm of \cite{Guan2022} which solves sparse generalized eigenvalue problem with $\ell_1$ penalty. The number of directions is estimated using 5-fold cross-validation.

We consider multiple $(n,p)$ pairs: 
$(100,200)$, $(500, 200)$, $(100,1000)$, and $(200,1000)$. To evaluate the prediction performance, we simulate an independent test set $ (Y^*_i(t), X^*_i) $ of size $ n^*=5000 $. Let $ \widehat \V $ be the optimal directions our procedure finds and $ \widehat{\gamma}(t) $ denotes the supervised principal component regression estimate obtained by regressing $ \mathbb{Y}(t) $ on the supervised principal components $\mathbb{X}\widehat \V $.  We measure the performance in terms of the optimal dimension $\widehat{K}$ and prediction error, which is defined as $ \sum_{i=1}^{n^*}\int_{0}^1 \{Y^*_i(t)-(X^*_i)^{\T} \widehat \V \widehat{\gamma}(t)\}^2 dt / n^{*}$ 
through 100 Monte Carlo simulations. For setting 1, since we know the ground truth of the subspace, 
the loss of subspace estimation is also  assessed using the Frobenius norm $\|\widehat{\Pi}-\Pi\|_{\mathrm{F}}^{2}$,
where $\Pi = \V^*(\V^*)^{\T}$ and $\widehat{\Pi}=\widehat{\V}_{(K_0)}\widehat{\V}_{(K_0)}^{\T}$ with $\widehat{\V}_{(K_0)}=(\widehat{\V}_{1},\ldots,\widehat{\V}_{K_0})$. 
Simulation results for setting 1 are summarized in Table \ref{table1}. We present the results under setting 2 in Table S3 of the supplementary material. 

\begin{table}[!ht]
  \begin{center}
    \caption{\label{table1}  Simulation results for setting 1: ``SPCR" represents the proposal \eqref{optV_penalize}; ``SCPR-no penalty" is similar to ``SPCR" except that it does not have the $\ell_1$ penalty on $\V$; 
``PGD-GEV" is the proximal gradient descent algorithm proposed by \cite{Guan2022}; 
``riﬂe-GEV" is the truncated Rayleigh flow method of \cite{tan2018sparse};
``UPCR" refers to the conventional principal component regression; ``Superpc" represents the supervised principal component method \citep{Bair2006} and ``CCA" is the canonical correlation analysis. Standard errors are presented in the bracket if applicable. For simplicity, we write $\text{SE}=0.00$ if $\text{SE}<0.005$. ``$-$" stands for inapplicable. All results are based on 100 Monte Carlo runs}
    \resizebox{\columnwidth}{!}{%
    \begin{tabular}{crccccccc}
    \toprule
     & & SPCR & SPCR-no penalty & PGD-GEV & rifle-GEV &  UPCR & Superpc &CCA  \\
\midrule \midrule
&&\multicolumn{6}{c}{$(n,p)=(100, 200)$} & \\
\midrule
Prediction Error && 1.46(0.03) & 20.84(0.16)  & 1.29(0.00) & 9.15(0.01)  & 24.90(0.13)& 2.37(0.16) & 1.48(0.07)\\ 
$\|\widehat{\Pi}-\Pi\|_{\mathrm{F}}^{2}$& &  0.02(0.00)&0.02(0.01)&0.01(0.00)& 0.73(0.00)& $-$ & $-$ & $-$\\
$\widehat{K}$  && 3.07(0.03)&5.69(0.24)&3.00(0.00)& 1.00(0.00)&18.36(0.34)&7.99(0.20)&4.70(0.06)\\
\midrule 
&&\multicolumn{6}{c}{$(n,p)=(500, 200)$} & \\
\midrule
Prediction Error &&  1.08(0.00)&9.61(0.06)
&1.05(0.00)&9.13(0.01)&24.33(0.12)&1.48(0.04)
&1.47(0.03) \\
$\|\widehat{\Pi}-\Pi\|_{\mathrm{F}}^{2}$& &  0.02(0.00)
&0.02(0.00)&0.00(0.00) &0.73(0.00)& $-$ & $-$ & $-$\\
$\widehat{K}$  &&  3.70(0.05)&5.40(0.25)&
3.09(0.06) &1.00(0.00)&19.87(0.04)&4.56(0.16)&
4.94(0.03)\\
\midrule 
&&\multicolumn{6}{c}{$(n,p)=(100, 1000)$} & \\
\midrule
Prediction Error && 1.50(0.06)&32.30(0.99)
&1.33(0.00)&9.21(0.02)&28.21(0.05)&6.19(0.50)
&4.22(0.10)  \\
$\|\widehat{\Pi}-\Pi\|_{\mathrm{F}}^{2}$&& 0.02(0.00)
&0.07(0.02)&0.01(0.00)&0.73(0.00)&  $-$ & $-$ & $-$   \\
$\widehat{K}$  && 3.99(0.08)&2.02(0.23)
&3.00(0.00)& 1.00(0.00)
&16.41(0.52)&6.21(0.36)&4.53(0.08) \\
\midrule 
&&\multicolumn{6}{c}{$(n,p)=(200, 1000)$} & \\
\midrule
Prediction Error && 1.13(0.01)
&25.07(0.08)&1.06(0.00)& 9.13(0.01)
&28.26(0.05)&1.78(0.06)&2.90(0.05) \\
$\|\widehat{\Pi}-\Pi\|_{\mathrm{F}}^{2}$&& 0.01(0.00)
&0.00(0.00)&0.00(0.00) & 0.73(0.00)& $-$ & $-$ & $-$ \\
$\widehat{K}$  &&4.52(0.06)&4.20(0.27)&
3.01(0.01)&1.00(0.00)&16.20(0.55)&7.33(0.16)&
4.88(0.04)   \\
\bottomrule
\end{tabular}%
}
\end{center}
\end{table}
Table \ref{table1} suggests that under setting 1 where $\Sigma_{xy}$ is low rank and $\V^*$ is sparse, ``PGD-GEV" has slightly better performance than our method ``SPCR" in terms of prediction and recovery 
of the given true subspace. Moreover, they both outperform the other methods. 
From Table S3 in the supplementary, we can see that the superior performance of ``SPCR" and ``PGD-GEV" are further demonstrated under setting 2 where the two key assumptions -- low rank and sparsity are violated. Our SPCR admits lower prediction error when $(n,p)=(200, 1000)$, while in other cases ``PGD-GEV" works better.

The outperformance of ``PGD-GEV" is attributed to directly solving the original sparse generalized eigenvalue problem \eqref{optW_sparse}. In the revision stage of the paper, one referee pointed out a very recent work of \cite{Guan2022}. For practitioners interested in an exact solution, we recommend the sequential procedure in \citet{Guan2022} as it performs slightly better than our approach for most cases. Regarding computation time, our algorithm is computationally faster, particularly for large $ p $, as we employ a convex relaxation of \eqref{optW_sparse} and can solve multiple directions simultaneously. Tables S16 and S17 in the supplement showcase our method's computational efficiency across various data generation settings by displaying the average computation time.

\subsection{Functional responses with measurement error}\label{merror}
To further evaluate the performance of the proposed method, we design additional numerical experiments where the functional responses are densely observed with measurement errors. 
In the simulation, we set the time points $t_{il}$ to be 1000 equally spaced time points between 0 and 1, i.e., $ t_{il}=t_{l} = l/1000$ for $ i=1, \ldots, n$ and $ l=1, \ldots, 1000$. The data $\{(X_i, Y_i(t))\}_{i=1}^n$ are generated in the same way as setting 1. The contaminated responses follow $ W_{il}= Y_i(t_{il})+ \epsilon_{il}$ for $i=1, \ldots, n$ and $l=1, \ldots, 1000$, where 
$\epsilon_{il}$'s are independently generated from  $N(0, v^2)$. Here we consider $v^2 = 1, 5, 10$. 
For each individual curve, applying smoothing spline based on the noisy observations $\{W_{il}\}_{l=1}^{1000}$ gives smoothed curve $\tilde{Y}_i(t)$ and their realization $\tilde{Y}_i(t_{l})$ at the grid points. Then we apply the proposed approach and other competing methods to $\{\tilde{Y}_i(t_{l})\}_{l=1}^{1000}$. Moreover, we also compare with the procedures: directly using the error contaminated data $\{W_{il}\}_{l=1}^{1000}$ without smoothing; and taking $\{Y_i(t_{l})\}_{l=1}^{1000}$ as observed responses by pretending that we know the underlying function a priori, referred to as oracle estimation. 

Tables S4--S9 in the supplement present the simulation results with error variance $v^2= 1,5,10$, respectively, under setting 1. Similar analysis is also performed under setting 2 and the results are reported in Tables S10--S15 of the supplement. 
Empirical evidence suggests that the smoothing step can improve the estimation and prediction performance within each method, and the result is close to the oracle estimate. 
Moreover, similar findings to the first simulation study can be obtained: the proposed "SPCR" and the proximal descent algorithm "PGD-GEV" exhibit comparable performance and outperform other methods in various settings,  irrespective of the smoothing.

To provide further insight into the sensitivity of the proposed method to the assumption of banded covariance, we carry out additional numerical studies where the covariace matrix $\Sigma_x$ is non-bandable. 
Specifically, the performance of SPCR is evaluated in three scenarios under both setting 1 and setting 2: $\Sigma_x$ follows a long-range dependence structure as described in \cite{BICKEL2008}; $\Sigma_x$
has randomly assigned non-zero off-diagonal elements; $\Sigma_x$ is block diagonal in which one block is a dense matrix with equal off-diagonal entries. 
The simulation results are reported in Tables S19--S21 of the supplement, respectively, which suggest that the proposed procedure still performs favorably in most settings, especially when the dimensionality increases or the assumptions of low rankness and sparsity are violated. This indicates that ``SPCR" is fairly robust to the banded assumption. 
See detailed discussion in Section S9 of the supplement.

\section{Real Data Application}\label{s5}
To explore the potential benefit of our method, we apply the proposed method to the cortical surface emotion task-related functional magnetic resonance imaging (fMRI) data from Human Connectome Project Dataset (https://www.humanconnectome.org/). We include the data usage acknowledgment in Section S1 of the supplementary material. We use the 900 Subjects release that includes behavioral and 3T MR imaging data of 970 healthy adult participants collected between 2012 and 2015. Our analysis includes 
805 participants with the cortical surface emotion task-evoked fMRI data available.

This emotion task is similar to the one developed by Hariri and his colleagues \citep{Hariri2002}. Participants are presented with blocks of trials that either ask them to decide which of the two faces presented at the bottom of the screen matches the face at the top of the screen or which of the two shapes presented at the bottom of the screen matches the shape at the top of the screen. The faces have either angry or fearful expressions. Trials are presented in blocks of 6 trials of the same task (face or shape), with the stimulus presented for 2 seconds and a 1 second inter-trial interval. Each block is preceded by a 3-second task cue (shape or face), so each block is 21 seconds, including the cue. Each of the two runs includes three face blocks and three shape blocks. For each subject, the number of frames per run of the emotion task is 176, with a run duration of 2.16 minutes.

We use the ``Desikan-Killiany'' atlas \citep{Desikan2006} to divide the brain into $68$ regions of interest. In this analysis, we focus on the association between clinical variables and BOLD signals in the left isthmus of the cingulate gyrus region. 
As a major component of the limbic system \citep{Broca1878}, the cingulate gyrus is involved in processing emotions \citep{MacLean1990} and behavior regulation. Damage to the cingulate gyrus may result in cognitive, emotional, and behavioral disorders. For each subject $i$, at each fixed time point, we average the blood oxygenation level-dependent signals of all pixels in the left isthmus of the cingulate gyrus, which results in a functional curve $ Y_i(t) $ observed on 176 equally spaced time points. We consider $p=280$ covariates as our predictor $ X $. We group these covariates into different categories: 
demographic, language, emotion, motor, and free surfer brain summary statistics so that the covariance of $X$ has a bandable structure. 
We apply the proposed method to the data and obtain two supervised principal components. Eleven of the nonzero elements in the first supervised principal component are in the emotion category. Specifically, two are from the Penn Emotion Recognition Test, namely ``Number of Correct Anger Identifications" and ``Number of Correct Fear Identifications." The others are behavior scores for NIH Toolbox Anger Affect Survey, NIH Toolbox Anger Hostility Survey, NIH Toolbox Anger Physical Aggression Survey, NIH Toolbox Fear Affect Survey, NIH Toolbox Fear Somatic Arousal Survey, NIH Toolbox Meaning, and Purpose Survey, NIH Toolbox Friendship Survey, NIH Toolbox Perceived Hostility Survey, and NIH Toolbox Perceived Rejection Survey. We plot the estimated coefficient functions $\widehat{\gamma}_1(t)$ and $\widehat{\gamma}_2(t)$ corresponding to the selected two supervised principal components in Figure S1-(a) and S1-(b), respectively, of the supplement. To evaluate the prediction performance, 
we randomly pick $\floor{2n/3}$ subjects as training data and the remaining as test data. We calculate the average prediction error by repeating this step 100 times. We also compare the competing methods considered in the simulations using the same data sets as SPCR. We report the prediction errors in Table \ref{tbl-realData}, which shows that our procedure achieves the best prediction performance and lends further support to the advantage of the proposed method. 

\begin{table}[!ht]
\centering
\caption{Average prediction errors based on 100 random splits. Standard errors are presented in the bracket
}
\resizebox{0.9\textwidth}{!}{%
\begin{tabular}{>{\rowmac}c>{\rowmac}c>{\rowmac}c>{\rowmac}c>{\rowmac}c>{\rowmac}c>{\rowmac}c>{\rowmac}c<{\clearrow}}
\toprule
Method  & SPCR & SPCR-A & PGD-GEV & SPCR-no penalty & UPCR & Superpc& CCA  \\ 
\midrule
Prediction Error & 1.81(0.002)  & 1.84(0.002) & 1.88 (0.001) & 2.22(0.002) & 1.85(0.002) & 1.83(0.001) & 1.84(0.001) \\
\bottomrule
\end{tabular}%
}
\label{tbl-realData}
\end{table}

\section{Discussion}\label{s6}
In this paper, we propose a novel supervised principal component regression approach for a linear model with functional response and high dimensional covariates. We develop an efficient estimation procedure by reformulating the nonconvex sparse generalized Rayleigh quotients problem into a convex optimization and provided theoretical guarantee for the reformulation. We also derive an non-asymptotic error bound for the estimated supervised principal component directions. Our method is evaluated through simulation studies and a real-world application of the Human Connectome Project fMRI data. There are several potential extensions of our method. First, it worth further investigating how to develop statistical inference procedures such as hypothesis testing and constructing confidence regions. Second, we only consider the case when the response is a single function. Motivated by our real data application, it would be interesting to study the case when we have multiple functional responses \citep{ding2021multivariate}.  We assume a linear relationship between the response and the predictor. It remains open how to obtain the supervised principal component directions when the true relationship is nonlinear or the model is misspecified. 
Finally, our theoretical results are restricted to the banded covariance matrix class for the covariates. Extending the method to sparse or low-rank covariance matrices is an interesting future direction. 
These extensions are beyond the scope of this paper, and we leave them for future research.

\section*{Supplementary Materials}
The supplement contains acknowledgment of data usage, all technical details and additional simulations, together with code for reproducing the numerical results.

%\section*{Acknowledgement}
%The authors thank the Editor, Associate Editor and referees for their constructive comments and suggestions. Sun's research is partially supported by the Natural Sciences and Engineering Research Council of Canada, a New Frontiers in Research Fund NFRFE-2019-00603 and a Data Sciences Institute Catalyst Grant. 
%Kong's research was partially supported by the Natural Sciences and Engineering Research Council of Canada.

%\section*{Disclosure Statement}
%The authors report there are no competing interests to declare.

\bibliographystyle{agsm}

\bibliography{SPCRbib1.bib}

\end{document}